\documentstyle[prl,aps,twocolumn,graphicx]{revtex}

\begin{document}

\newcommand\xmas{\mbox{$x_{\rm mas}$}}
\newcommand\rmas{\mbox{$r_{\rm mas}$}}
\newcommand\Xmas{\mbox{$X_{\rm mas}$}}
\newcommand\Hinf{\mbox{${\rm H}_{\rm inf}$}}
\newcommand\Omegainf{\mbox{$\Omega_{\rm inf}$}}
\newcommand\Omegabg{\mbox{$\Omega_{\rm bg}$}}
\newcommand\Hbg{\mbox{${\rm H}_{\rm bg}$}}
\newcommand\rhobg{\mbox{$\rho_{\rm bg}$}}
\newcommand\rhorbg{\mbox{$\rho^{\rm bg}_{\rm r}$}}
\newcommand\rhoinf{\mbox{$\rho_{\rm inf}$}}
\newcommand\kinf{\mbox{$k_{\rm inf}$}}
\newcommand\kbg{\mbox{$k_{\rm bg}$}}
\newcommand\ainf{\mbox{$a_{\rm inf}$}}
\newcommand\abg{\mbox{$a_{\rm bg}$}}
\newcommand\xpbg{\mbox{$x_{\rm p}^{\rm bg}$}}
\newcommand\xpinf{\mbox{$x_{\rm p}^{\rm inf}$}}
\newcommand\xmasbg{\mbox{$x_{\rm mas}^{\rm bg}$}}
\newcommand\xmasinf{\mbox{$x_{\rm mas}^{\rm inf}$}}
\newcommand\thetainf{\mbox{$\theta^{\rm inf}$}}
\newcommand\thetabg{\mbox{$\theta^{\rm bg}$}}
\newcommand\xHbg{\mbox{$1/{\rm H}_{\rm bg}$}}
\newcommand\xHinf{\mbox{$1/{\rm H}_{\rm inf}$}}


\title{Inflationary Initial Conditions Consistent with Causality}

\author{Arjun Berera
\thanks{email:ab@ph.ed.ac.uk, PPARC Advanced Fellow}
and 
Christopher Gordon
\thanks{email:christopher.gordon@port.ac.uk}
}

\address{ 
$^*$Department of Physics and Astronomy, 
University of Edinburgh, 
Edinburgh, 
EH9 3JZ, U.K. \\
$^{\dag}$ Relativity and Cosmology Group, Division
of Mathematics and Statistics, \\ Portsmouth University,
Portsmouth~PO1~2EG, U.K.}


\maketitle

\begin{abstract}
  The initial condition problem of inflation is examined from the
  perspective of both spacetime embedding and scalar field dynamics.
  The spacetime embedding problem is solved for arbitrary initial
  spatial curvature $\Omega$, which generalizes previous works that
  primarily treat the flat case $\Omega=1$. Scalar field dynamics that
  is consistent with the embedding constraints are examined, with the
  additional treatment of damping effects. The effects of
  inhomogeneities on the embedding problem also are considered.  A
  category of initial conditions are identified that are
not acausal
and can develop into an inflationary regime. 

\end{abstract}

\vskip1pc


\vskip1pc

In press Physical Review D 2001. 

hep-ph/0010280.


\section{Introduction}
\label{introduction}
Inflation is the foremost idea for explaining the large scale
homogeneity and near flatness of the universe, which are the two main
unexplained observational features in the standard hot
big-bang model.  The large scale homogeneity or horizon problem
amounts to the fact that under hot big-bang evolution, due to a
decelerating scale factor ${\ddot a} < 0$, sufficiently separated
regions of the present day observable universe would never have been in
causal contact.  Nevertheless, it appears the
universe we observe looks very much the same, and in particular very
smooth, in all directions.  This fact is best seen in the cosmic microwave
background radiation (CMB) which has  temperature fluctuations of
only one part in $10^5$ when measured from any direction in the sky
\cite{Bennett:1996ce}.

The inflation solution to this horizon problem is to picture the
universe during an early epoch to undergo an accelerated expansion,
${\ddot a} > 0$.  Such expansion can take an initially small causally
connected patch and enlarge it to a size that comfortably encompasses
our present day observed universe, thereby solving the horizon problem.
To realize inflation, the equation of state within the inflationary
patch must be of a very special form, possessing negative pressure.
The potential energy of a scalar field has an equation of state that
satisfies this requirement. This fact has been a key link towards a
dynamical realization of inflation and thereby has further motivated
the inflation solution.

As inflation is meant to solve the horizon problem, it is important
that it does not require acausal initial conditions. 
In particular, 
the picture of inflation considered here
is for the universe to emerge from an initial singularity
and then 
enter into a hot big-bang radiation dominated evolution.  At some
time $t_i$ after the initial singularity, the conditions appropriate
for inflation should occur within a small patch that is contained
within the causal horizon at that time.  
Chaotic inflation \cite{ci,Linde:1990nc}  does not fall into this picture
as there inflation 
is thought to start at the Planck epoch with homogeneity assumed on
the Planck scale.
To realize the picture of a
``local'' inflation, two requirements must be satisfied.  {}First a
physically sensible embedding must be demonstrated of the inflating
patch immersed within a non-inflating 
background \cite{embedd,Vachaspati:2000dy,Trodden:1999wc}.  
By embedding we mean matching the inflationary space time with the
background space time at the boundary of the patch.
Second, it must be
shown that
for $t> t_i$, the patch is dynamically stable to sustain
inflation \cite{piran181,kb,kb90,gpprl,gold43,other,Goldwirth:1992rj}.
{}For example, large fluctuations, which conceivably could
enter the inflating patch at a maximum rate limited only by causality,
should not destroy the inflationary conditions within the patch.

Recently, a convenient methodology has been developed
in \cite{Vachaspati:2000dy,Trodden:1999wc} for analyzing
the embedding problem, based
on the null Raychaudhuri equation \cite{ray,Wald:1984rg}.  In Sec. \ref{embedding_conditions}
the flat spacetime ($\Omega=1$)
formulation in \cite{Vachaspati:2000dy} 
is generalized to arbitrary spacetime curvature.
Then, an alternative solution from \cite{Vachaspati:2000dy}
to the embedding problem will be identified, which is especially
attractive for an 
inflating patch with an open geometry.
In Sec. \ref{dynamic_conditions}, initial conditions
for scalar field dynamics are presented, 
which are consistent with causality and our
embedding solution and which evolve into
successful supercooled or warm inflationary regimes.
In Sec. \ref{dynamical_effects_on_embedding} we combine the results in
Sec. \ref{embedding_conditions}  and Sec. \ref{dynamic_conditions} to evaluate
the effect of inhomogeneities on the embedding problem.
Finally, Sec. \ref{conclusion} presents our conclusions.

\section{Embedding Conditions}
\label{embedding_conditions}
The embedding of an inflationary
patch within a background space time generally is
regarded as unacceptable if there are any negative energy
regions. This requirement often is referred to as
the weak energy condition.  
The null Raychaudhuri equation is a useful diagnostic
tool for determining the validity of this condition.  This
equation determines the evolution
of the divergence $\theta \equiv \nabla^{\alpha}N_{\alpha}$
of the null ray vector $N^\alpha$ 
in a spacetime
with an arbitrary, and {\em not necessarily homogeneous},
energy density distribution.  
In order for the weak energy
condition to be valid, this equation
implies that for a null geodesic the condition 
\cite{Vachaspati:2000dy,Wald:1984rg}
\begin{equation}
\frac{d \theta}{ds} \leq 0
\label{dtheta}
\end{equation}
must be satisfied,
where $s$ is an affine parameter along the
null geodesic.  

{}For application of Eq.\ (\ref{dtheta}) to the inflation
embedding problem, the concept of anti-trapped
and normal regions should be understood.
Consider a sphere centered on a comoving observer.
If the space time were not expanding, photons emitted from the surface
of the sphere, radially towards the observer, would converge at
the observer. However, the expansion tends to work against the bundle
of rays converging to a point. If the expansion is rapid enough, the
bundle of rays will have diverging trajectories and then the spherical
surface from which the rays originated is
said to be an {\em anti-trapped\/} surface \cite{Wald:1984rg}.  
The spherical surface 
with a radius $\xmas$ is known as
the {\em minimally anti-trapped surface} (MAS) if any sphere with a
larger radius is anti-trapped.
{}For inwardly directed null rays, the divergence, $\theta$, will
be positive in anti-trapped regions of space.
On the other hand, 
if $\theta$ is negative for
inwardly directed null rays and positive for outwardly directed null
rays, the region is called {\it normal}.

These definitions are convenient  when
analyzing the weak energy condition based on Eq.\ (\ref{dtheta}).
{}For example, one can
immediately conclude \cite{Vachaspati:2000dy} that if an outer normal region
bounds an inner anti-trapped region in a not necessarily homogeneous,
spherically symmetric space time, the
weak energy condition would be violated, thus implying negative energy
is required at the boundary of such a configuration.

{}For the inflation problem, the inner
region is modeled as the putative inflation
patch (INF) and it is immersed within a outer background (BG)
expanding spacetime region.
As a first approximation, both the INF and BG regions are
characterized by FRW metrics of the form
\begin{equation}
ds^2 = -dt^2 + a^2 \left( \frac{dr^2}{1-\kappa r^2} + r^2 d\Omega^2 \right),
\label{frw}   
\end{equation}
where $\kappa=-1$ for an open geometry, 0 for a flat geometry and 1 for a
closed geometry. 
The effect of inhomogeneities in the BG will be considered in
Sec. \ref{dynamical_effects_on_embedding}. The INF patch should be homogeneous.
In what follows, $\theta$ and $\xmas$ will be computed for
a spacetime characterized by the metric Eq.\ (\ref{frw}).
The results below are applicable for both the
INF and BG regions, given that the appropriate parameters are used
in Eq.\ (\ref{frw}) for the two different spacetime regions.

Proceeding with this calculation,
the determinant of the metric Eq.\ (\ref{frw}) is
\begin{equation}
  \label{eq:determinant}
  g = |\mbox{det}(g_{\mu\nu})| = a ^ 6 r ^ 4 (1 -
\kappa r^2)^{-1}\sin^2\psi ,
\end{equation}
where $\psi$ is the polar angle.
An incoming light ray (null vector) is given by
\begin{equation}
  \label{eq:null vector}
  N^\alpha
 = \frac{1}{a}\left[\delta^{\alpha}_{0} - a ^ {-1} (1 - \kappa r^2)^
  {1/2}\delta^\alpha_1\right] ,
\end{equation}
with the divergence of the null rays being
\begin{equation}
  \label{eq:divergence}
  \theta
 = \frac{1}{\sqrt{g}} (\sqrt{g}N^\alpha)_{,\alpha}.
\end{equation}
Substituting equations (\ref{eq:determinant}) and (\ref{eq:null vector})
into (\ref{eq:divergence}) gives
\begin{equation}
  \label{eq:soln1}
  \theta
= \frac{2}{a} \left( {\rm H} - \frac{(1 - \kappa r^2)^{1/2}}{ar} \right)
\label{thetaio}
\end{equation}
with ${\rm H} \equiv {\dot a}/a$ the Hubble parameter.
The above recovers equation (10) 
of \cite{Vachaspati:2000dy} for $\kappa = 0$.

The physical distance is given by
\begin{eqnarray}
  x &=& a \int dr (1-\kappa r^2)^{-1/2} \\ &=& a\, \mbox{arcsinh(r)}, \quad
    \mbox{for $\kappa = -1$}
  \label{eq:distance}
\end{eqnarray}
The MAS comoving distance, $\rmas$, must
give a zero divergence and so from Eq.\ (\ref{eq:soln1})
with, for example, $\kappa=-1$
\begin{equation}
  \label{eq:condition}
  {\rm H} - \frac{{(1 + \rmas^2)}^{1/2}}{a\rmas} = 0,
\end{equation}
from which we get, using equation (\ref{eq:distance}),
\begin{equation}
  \label{eq:open_cond}
  \xmas = a\, \mbox{arcsinh}\left[({\rm H}^2a^2 - 1)^{-1/2}\right].
\end{equation}
{}From the Friedman equation:
\begin{equation}
  \label{eq:curv-density}
  {\rm H}^2 a^2 = \frac{\kappa}{\Omega - 1}
\end{equation}
Substituting (\ref{eq:curv-density}) into (\ref{eq:open_cond})
gives the solution for $\xmas$ as a function of $\Omega < 1$.
A similar equation can be derived for the closed case $(1 < \Omega <
2)$ where an upper limit is needed on $\Omega$ to ensure 
that $1/{\rm H}$
is smaller than the radius of the Universe.  
Therefore, the general MAS size is given by
\begin{equation}
  \label{mascond}
  \xmas(t) = \frac{1}{\rm H}\left\{
\begin{array}{ll}
\frac{1}{(1-\Omega)^{1/2}} \mbox{arcsinh} \left(\sqrt{\frac{1 -
      \Omega}{\Omega}} \right), & 0 < \Omega < 1 \\ 1, & \Omega = 1 \\
      \frac{1}{(\Omega - 1)^{1/2}} \mbox{arcsin}
      \left(\sqrt{\frac{\Omega - 1}{\Omega}} \right), & 1< \Omega < 2 
\end{array}
\right.
\end{equation}
\begin{figure}[htbp]
   \begin{center}
         \includegraphics[angle=-90]{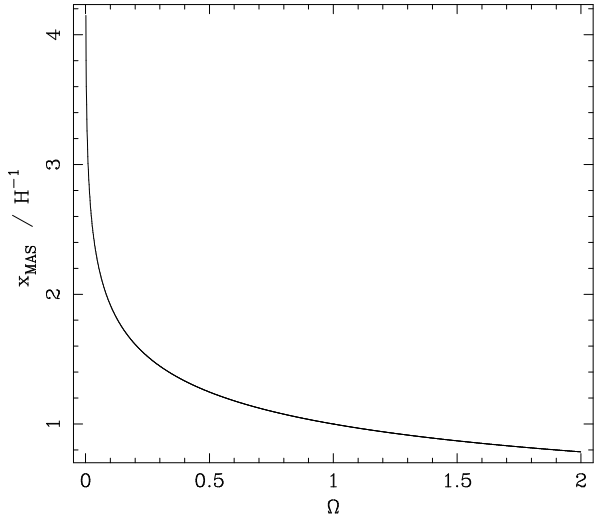}
     \caption{A plot of MAS size as function of the density.}
     \label{fig:xmas,Omega}
   \end{center}
\end{figure}
A plot of equation (\ref{mascond}) is given in Fig.
\ref{fig:xmas,Omega}. As can be seen, 
the size of the MAS becomes arbitrarily large relative
to the Hubble radius as $\Omega \rightarrow 0$.

The results for the $\xmas$ derived above are valid for either the
INF or BG region. In \cite{Vachaspati:2000dy}, the case $\Omega=1$ was
treated, and our calculation above is consistent with them in this
limit.  Before proceeding with our analysis for general $\Omega$, let
us review the results of \cite{Vachaspati:2000dy} for $\Omega=1$.  The
background spacetime in their work is pictured, as for us, as evolving
in a hot big bang regime.
{}For $\Omega=1$ and standard forms of matter,
e.g. radiation, non-relativitic matter, etc...,
the Hubble horizon sets the
the scale on which causal processes can take place.
For definiteness, if  we take the background causal horizon size as
$\xpbg = 1/\Hbg$, then by Eq.\ (\ref{mascond}) for $\Omega=1$, $\xpbg
=\xmasbg$.
\begin{figure}[htbp]
  \begin{center}
    \includegraphics[width=\hsize]{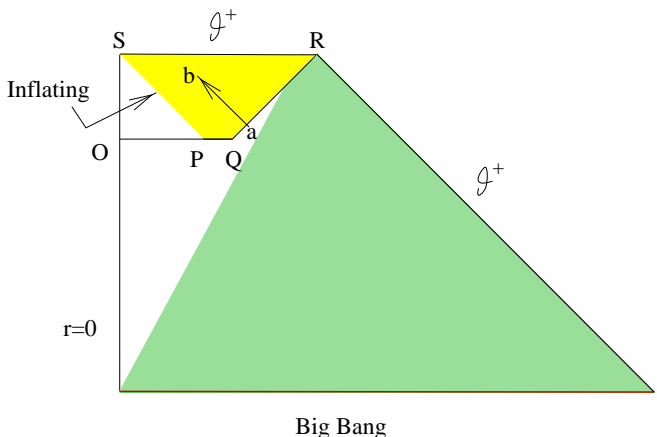}
    \caption{A Penrose diagram for local inflation (adapted from
\protect \cite{Vachaspati:2000dy}
). The arrow (ab) denotes a radially directed null-geodesic
      going from the normal background space time to the anti-trapped
      space time inside the patch. Shading represents anti-trapped
      regions.} 
    \label{fig:penrose}
  \end{center}
\end{figure}
Fig. \ref{fig:penrose} shows a Penrose diagram for an embedding that
violates the weak energy condition.  Here, at the beginning of
inflation, $\xpinf$ is represented by line OQ and $\xmasinf$ is
represented by line OP.  An inflating patch is pictured to develop
within some region inside the background (polygon OQRS in Fig.
\ref{fig:penrose}).  In the inflationary patch, they assume a
different Hubble parameter $\Hinf$ and assume the size of this patch
must be $\xpinf > 1/\Hinf$.  This assumption can be justified by
requiring the patch to be stable against perturbations and will be
discussed further in the next section.  By Eq.\ (\ref{mascond}) it
also implies $\xmasinf = 1/\Hinf$ for $\Omega=1$.  If the patch is to
be set up by causal processes then we would expect it to be smaller
than the background Hubble horizon, $\xpinf < \xHbg$.  These
conditions combined imply $\xmasbg = \xHbg > \xpinf > \xmasinf$.  As
such the region between $\xmasbg$ and $\xpinf$ is normal with respect
to the BG-region and the region between $\xpinf$ and $\xmasinf$ is
anti-trapped with respect to the inflating patch. Thus an in-going
null ray will go from a negative divergence region to a positive
divergence region.  Based on Eq.\ (\ref{dtheta}) this leads to a
violation of the weak energy condition.  Due to this fact, in
\cite{Vachaspati:2000dy} they conclude that the only way to avoid this
violation is to have $\xpinf > \xHbg$ which may be impossible or at
least very difficult to come about through causal processes.

Although this analysis assumed a homogeneous background, the general
result of \cite{Vachaspati:2000dy} was that if the inflationary patch
contained a MAS then it must be larger than the background MAS. One of
the main aims of this article is to further examine the implications
this has for a causal embedding of an inflationary patch.

Our first observation is that there is an embedding consistent with
both causality and the weak energy condition, in particular
\begin{equation}
\xpbg , \xmasinf > \xpinf\,.
\label{embed2}
\end{equation}
Since $\xpinf$ in Eq.\ (\ref{embed2}) is smaller than $\xpbg$,
it implies consistency with causality.
A region which is smaller than its MAS will have no anti-trapped surfaces.
  Thus
in our proposed embedding Eq.\ (\ref{embed2}), the INF-BG boundary
is between two normal regions. 
As such, provided                                                           
$\thetainf]_{\rm boundary}$ is sufficiently smaller than $\thetabg]_{\rm
boundary}$,  the weak energy condition is satisfied.
One interesting feature of the embedding Eq. (\ref{embed2})
is that for $\Omegainf < 1$ it is acceptable for
$\xpinf > 1/{\rm H}_{\rm inf}$.

As inflation proceeds, $\Omegainf \rightarrow 1$ and so $\xmasinf
\rightarrow \xHinf$. One of the main features of inflation is that 
modes of a matter perturbation  become larger than the Hubble radius
as the Universe expands.
It follows that eventually $\xpinf > \xHinf, \xmasinf$ will need to
occur.
This will not
cause  violation of the weak energy condition provided that $\xpinf >
\xmasbg$ at that time.

\begin{figure}[htbp]
  \begin{center}
    \includegraphics[width=\hsize]{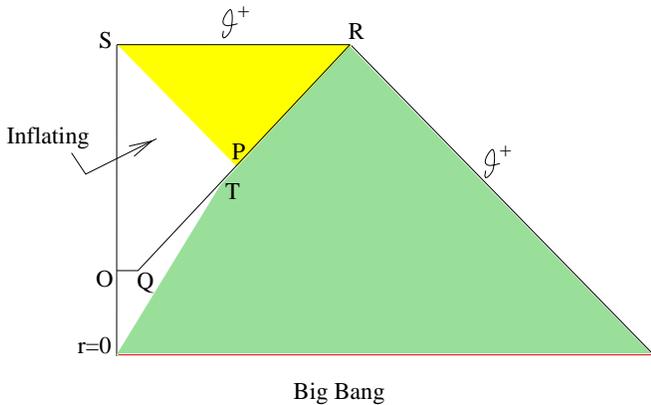}
    \caption{A Penrose diagram for local inflation with an embedding
      that does not violate the weak energy condition and is not
acausal.} 
    \label{fig:embed}
  \end{center}
\end{figure}

{}Fig. \ref{fig:embed} shows a Penrose diagram of the proposed
embedding.
At point T in the figure, $\xpinf=\xmasbg$ and at point P,
$\xpinf=\xmasinf$.
 As can be seen the anti-trapped region develops in the
patch only after the patch has become greater than the background MAS.

There is another embedding which does not violate the weak energy
condition, in this case for $\Omegabg > 1$, but 
it proves not to be as useful,
\begin{equation}
  \label{eq:closed_bg}
  \xmasinf, \xmasbg < \xpinf < \xHbg\,.
\end{equation}
This solution is restricted by the upper limit on $\Omegabg$
that gives a lower limit (from Eq.\ \ref{mascond}) of $\xmasbg \approx
0.8/ \Hbg$ for $\Omega=2$. 
As the causal 
scale $\xpbg$ 
can be only of the order of $\xHbg$, such a large inflationary
patch may still be acausal.

Additional information about embedding constraints can be
obtained through the Israel junction conditions \cite{Israel:1966rt}.
In \cite{bkt,Sakai:1994fu}
conditions were derived for the  embedding of one FRW spacetime
within another. The Israel junction conditions require the extrinsic
curvature on the background side of the boundary to be
smaller than the extrinsic curvature on the patch side of the 
boundary if the boundary surface energy density is to be positive. 
It can then be shown \cite{bkt,Sakai:1994fu} that if the 
energy density of the background $\rhobg$
is greater than that of the patch $\rhoinf$, the Israel junction
conditions can be satisfied for an inflationary patch of arbitrary size 
and geometry with any
background
geometry. They also show for $\rhobg < \rhoinf$, an embedding
consistent with the Israel junction conditions and $\xpinf < \xHbg$
requires
the background geometry to be closed with the geometry
of the inflationary patch arbitrary.

\section{Dynamic Conditions}
\label{dynamic_conditions}
An immediate concern with the proposed embedding Eq.\ (\ref{embed2})
regards the size of the inflationary patch.  As far as the embedding
conditions are concerned, the patch size simply must lie below the
$\xmas$ line in Fig. 1.  In particular, for $\Omegainf > 1$ the
inflationary patch size must be 
$< 1/\Hinf$.  
However, for
$\Omegainf < 1$, 
the inflationary patch size can be 
$ >
1/\Hinf$.  The primary question which remains is what the minimal
initial inflation 
patch
size can be in order to be dynamically stable for
inflation to commence.  {}For example, if the initial patch is too
small, then large fluctuations initially outside the patch could enter
inside at a rate sufficiently fast to destroy the inflationary
conditions.  In this section, the dynamic conditions necessary
for inflation will be examined for both the scalar field
and a background radiation component.

\subsection{Scalar Field}  
\label{scalar_field}
{}For scalar field driven inflation, the general dynamical conditions
necessary for a finite spatial patch to initiate and sustain inflation
have been considered in \cite{piran181,kb,kb90,gpprl,gold43,other} 
and a comprehensive review has
been given in \cite{Goldwirth:1992rj}. 
Here the requirements addressed in these
works, primarily the review \cite{Goldwirth:1992rj}, 
will be systematically examined
with emphasis on determining their implications for the
minimal initial inflationary patch size.  Our considerations 
of scalar field dynamics will generalize those in the 
above stated works, which focused on supercooled inflationary
dynamics \cite{siold,ni,ci}, 
in that here warm inflation dynamics \cite{wi} also will be
treated.

The classical evolution equation for the scalar inflaton
field has the general form
\begin{eqnarray}
& &{\ddot \phi} ({\bf x}, t)  + [3{\rm H}+\Gamma] {\dot \phi}({\bf x},t) 
\nonumber\\
& - &a^{-2} \nabla^2 \phi ({\bf x},t) + 
\frac{\delta V(\phi)}{\delta \phi({\bf x},t)} =0 ,
\label{phieom}
\end{eqnarray}
where $\Gamma$ is a dissipative coefficient, which represents
the effective interaction of the inflaton with other fields.
Here and below $\phi({\bf x},t)$ is a classical real number
which contains the ensemble average of the quantum and/or thermal
fluctuations.
In supercooled inflation, it is assumed the inflaton is isolated, in 
which case $\Gamma=0$. Thus no radiation is produced during inflation
and the universe inflates in a supercooled state.  On the other hand,
in warm inflation the inflaton interacts with
other fields, and the $\Gamma {\dot \phi}$ term is the simplest
representation of their effect on the evolution of the inflaton,
with generally $\Gamma > {\rm H}$.  Due to these interactions,
radiation is dissipated from the inflaton
system into the universe throughout the inflation period.
In general, $\Gamma$ can vary as a function of the inflaton field mode
to which it is associated, but here
$\Gamma$ is treated as a constant.

There are two types of potential $V(\phi)$ in Eq.\ (\ref{phieom})
that typically are treated in inflationary cosmology, a purely concave
potential of the generic form $V \sim m^2 \phi^2 + \lambda \phi^4$ 
and a double well potential such as $V \sim (\phi^2 - m^2)^2$.
Our discussion to follow will focus of the former type of potential
and at the end we will comment on the case of
the double well potential.  {}Furthermore, for most of our discussion
it will be adequate to consider the simplest case
of a quadratic potential, 
$V=\frac{1}{2} m^2 \phi^2$.
In this case,
going to Fourier space where we put the universe in a box
\begin{equation}
\phi({\bf x}, t) = \sum_{\bf k} \phi_{\bf k}(t) 
e^{i{\bf k} \cdot {\bf x}} ,
\end{equation}
Eq.\ (\ref{phieom}) becomes,
\begin{equation}
{\ddot \phi}_{\bf k} + [3{\rm H}+\Gamma] {\dot \phi}_{\bf k}(t)
+a^{-2} k^2 \phi_{\bf k}(t) + m^2 \phi_{\bf k}(t) =0 ,
\label{phikeom}
\end{equation}
where $k^2 \equiv |{\bf k}|^2$ with ${\bf k}$ the comoving
wave-vector and ${\bf k}/a(t)$ the corresponding physical wave-vector
at time $t$.

A necessary condition for inflation is that the zero mode
$\phi_{{\bf k} =0}$, must have a sufficiently large and
long-sustained amplitude
so that the potential energy $V=\frac{1}{2}m^2 \phi_0^2$ dominates the equation
of state of the universe, thereby driving inflation.
This leads to the familiar
slow-roll conditions, which require the curvature of the potential
to be sufficiently flat so that Eq.\ (\ref{phikeom}) 
for the zero mode becomes
first order in time
\begin{equation}
{\dot \phi}_0 = -\frac{dV(\phi)/d\phi_0}{3{\rm H}+\Gamma}.
\label{phik1eom}
\end{equation}

The dynamic initial condition problem is that the above requirements
should not be too special and in particular should not
violate causality.  The most acute initial condition
problem discussed in \cite{Goldwirth:1992rj} and related works is that generally
the initial inflaton field configuration will be very inhomogeneous,
thus considerable contribution of gradient energy ($(\nabla \phi)^2$)
should be present.  In its own right, the gradient term
for comoving mode ${\bf k}$ has energy density 
$\rho_{\nabla} = ({\bf k}^2/2a^2) \phi_{\bf k}^2$
and equation of state $p_{\nabla} = -\rho_{\nabla}/3$.
{}From the scale factor equation
\begin{equation}
\frac{\ddot a}{a} = -\frac{4 \pi G}{3}(\rho + 3p) ,
\label{scalefac}
\end{equation}
the effect of the gradient energy vanishes on the right hand
side. As such, any additional contribution from vacuum
energy still should drive inflation.  However,
excited modes with $k/a > 3{\rm H}+\Gamma$ not only will possess
gradient energy, but based on Eq.\ (\ref{phikeom})
are under-damped. As such, they also have a kinetic energy contribution
$\rho_{\dot \phi} \sim {\dot \phi}^2_{\bf k}$, which has
an equation of state $p_{\dot \phi} = \rho_{\dot \phi}$.
If the kinetic energy components of these modes dominates
the energy density in the universe, then from 
Eq.\ (\ref{scalefac}) inflation will cease to occur.

{}For the case of supercooled inflation, since $\Gamma=0$,
the simplest way to avoid
this problem is to require
that modes with $k/a \gtrsim {\rm H}$ initially
should not be excited. In other words, the inflaton field
initially should be smooth up to physical scales larger than
$\sim 1/{\rm H}$.  However, under general conditions, the causal
size of the pre-inflation patch  also will be of order
the Hubble radius $1/{\rm H}$. As such
this homogeneity requirement on the initial inflaton field
impinges on
being acausal,
since this condition
essentially requires initially smooth conditions up
to the causal scale.

To treat excited modes with $k/a > {\rm H}$
in the supercooled inflation case, the initial
condition dynamics are much more
complicated.  A simple analytical method applied to this
situation is the effective density approximation presented in
\cite{piran181,gold43} and reviewed 
in Sec. 7.2 of \cite{Goldwirth:1992rj}.  In this approach,
the inhomogeneities of the inflaton are treated in determining its
evolution, but the effect of these inhomogeneities on the
metric is only treated in the Friedmann equation through
homogeneous terms that represent the effective gradient and
kinetic energy densities as
\begin{eqnarray}
\left(\frac{\dot a}{a}\right)^2 \equiv {\rm H}^2 = 
 \frac{8\pi G}{3}
\left[\frac{1}{2} {\dot \phi}_0^2 + \frac{1}{2} m^2 \phi_0^2 \right.
+&& \nonumber\\ \left. \frac{1}{2} \sum_{\bf k}[{\dot \phi}_{\bf k}^2 + 
(\frac{{\bf k}^2}{a^2} +m^2) \phi_{\bf k}^2] +\rho_r\right]
-\frac{\kappa}{a^2}.&& 
\end{eqnarray}
{}For initial field configurations with sufficiently excited
$k/a > {\rm H}$ modes so that their
gradient energies dominate the equation of state of the
universe, the effective density approximation indicates
the Friedmann and inflaton evolution equations are highly coupled.
In particular, the Hubble parameter will be dominated by the
gradient energy term and this will act back on the inflaton
evolution through the $3{\rm H} {\dot \phi}$ term.  
{}Furthermore, the initially large gradient terms also
can in turn induce large kinetic energy in the modes. The outcome
found in \cite{piran181,gold43,Goldwirth:1992rj}
(see also \cite{kb90}) 
for this case is that the universe expansion
is non-inflationary.  On the other hand, for the excited
$k/a > {\rm H}$ modes with smaller amplitudes, their evolution
will be oscillatory and once again the universe expansion is
non-inflationary.  {}For either
of these two possibilities,
this approximation method finds that after an initial period
of detaining the universe in a non-inflationary
regime, the effect of these higher modes becomes
negligible. At this point the remaining vacuum energy dominates and
eventually inflation proceeds.

Numerical simulations \cite{gpprl,gold43}
which exactly treat the effects on the
metric due to the inhomogeneities in the inflaton field,
in fact, do not support this final conclusion of the effective
density approximation.  Instead, the simulations
find that once modes with $k/a > {\rm H}$ are sufficiently
excited, inflation generally does not occur or
is highly suppressed.  Although the final conclusion
of the effective density approximation was incorrect, it at least
indicated that there is nontrivial interplay between the
inflaton and metric evolution equations 
once significant short distance inhomogeneities
are present in the inflaton field. 

In contrast to the supercooled case, once $\Gamma > {\rm H}$, the
effective density approximation indicates that for modes
with $\Gamma > k/a > {\rm H}$, there is almost no
feedback between the inflaton and Friedmann equations.
The evolution equation for these inflaton modes is
\begin{equation}
\Gamma \dot \phi_{\bf k}(t) = \left[\frac{{\bf k}^2}{a^2}+
m^2 \right] \phi_{\bf k}(t),
\end{equation}
and in particular 
is
independent of ${\rm H}$.  This in turn implies all these modes
are over-damped. As such, they only contribute gradient energy
to the equation of state of the universe, which alone is
ineffective in preventing inflation.  Thus up to the predictions
of the effective density approximation, in contrast to the
supercooled case,  in this warm inflation case modes with
$k/a < \Gamma$ show little coupling between the inflaton and
Friedmann equations. 
It is evident that dynamics with a $\Gamma {\dot \phi}$
damping term should have qualitative differences for the
initial condition problem compared to the case with $\Gamma=0$.
In particular, 
modes
of physical wavelength smaller that the Hubble radius
$1/{\rm H}$ but larger than $1/\Gamma$ could still be substantially
excited without destroying entrance into inflation.
This implies inflationary patch sizes smaller than the Hubble
radius may be dynamically stable in sustaining
inflation and based on the results of Sec. \ref{embedding_conditions}, 
they also
provide consistent spacetime embeddings, especially for 
$\Omegainf < 1$. 
{}Furthermore, studies of warm 
inflation \cite{wi,Berera:1997fm,taylor2000},
including the first principles quantum field theory
model \cite{wifp}, generally find
that to obtain adequate inflationary e-folds,
$N_e \stackrel{>}{\sim} 60$, it requires $\Gamma \approx N_em^2/{\rm H}$
with $m \gtrsim {\rm H}$ so that
$\Gamma \gtrsim N_e {\rm H}$ and thus $1/\Gamma \ll 1/{\rm H}$.
As such, for warm inflationary conditions this simple
analysis suggests that the smoothness requirement on the
initial inflationary patch is at scales much smaller than the
Hubble radius $1/{\rm H}$, which therefore imposes no
violation of causality.

These considerations can be extended to the more complicated situation
where the inflaton evolution equation has nonlinear terms,
such as the $\phi^4$ interaction term.  {}For supercooled inflation,
the effect of such interactions has been considered from the
perspective of mode mixing for some special cases in \cite{kb,kb90}
and through computer simulations of scalar fields in one spatial
dimension in \cite{gpprl,gold43}.  Up to the scope of these works,
their analysis concludes that nonlinear interactions do not present
any additional complications to the initial condition problem.
{}For the warm inflation case, this conclusion can be stated
in more general terms.  In particular, the dissipative coefficient 
$\Gamma$  completely damps the evolution, thus suppressing mode mixing,
of all modes with $k/a < \Gamma$.  As such, initially excited
modes with $k/a < \Gamma$ will evolve independently within
the time scales relevant to the initial period when inflation begins.
The only effect of excited modes with $k/a < \Gamma$ is to
contribute gradient energy to the equation of state of the universe,
and this alone is ineffective in suppressing inflation.
Therefore, provided the inflaton field also initially
maintains some vacuum energy, no aspect of the inflaton's dynamics
in the early period acts to circumvent entrance into the inflation 
phase.  

Although the above discussion addressed the 
main impediment in scalar field dynamics that
can prevent inflation,
there are a few smaller concerns worth
mentioning here.
One detail treated in \cite{Goldwirth:1992rj} is
inflaton initial conditions with large kinetic energy.
{}For $\Gamma=0$, it is shown in \cite{Goldwirth:1992rj} that this problem is 
self-correcting, and so should not pose a major barrier
in entering into the inflation phase.  {}For $\Gamma > {\rm H}$, the severity
of this problem diminishes further since this damping
term is an additional effect that helps suppress the
initial kinetic energy.  Another secondary detail is that above
we only treated concave potentials.  {}For double well potentials,
$V \sim (\phi^2 - m^2)^2$, as in the case of new 
inflation \cite{ni},
inflation requires the field to be well localized 
at the top of the potential hill $\phi =0$ and almost at
rest.  These requirements are necessary, otherwise as found
in various studies \cite{niearly}, the field can easily break up
into several small domains with randomly varying signs
of the amplitude.  {}For the supercooled case, new inflation,
the basic conclusion about the initial
conditions  required for inflation has been that they
are not very robust \cite{niearly,niic,Goldwirth:1992rj}.  The inclusion
of a dissipative term $\Gamma > {\rm H}$ will not necessarily improve this
situation.  On the one hand, such a term helps
considerably to damp kinetic energy, thus allowing
the inflaton to remain at the top of the hill longer
and drive inflation.  On the other hand, if such a damping term
arises too early before inflation is to begin,  since it freezes
the evolution of excited inflaton modes, it could prevent
the initial inflaton field configuration 
from equilibriating to $\phi=0$.

{}Finally in the above discussion $\Gamma \ne 0$ generally
represented warm inflation dynamics, although even for supercooled
inflation, if radiation is present at the onset of inflation,
such as in the new \cite{ni} and thermal \cite{ti}
inflation pictures, the effective evolution of the
inflaton could have a damping term of the form $\Gamma {\dot \phi}$.
Since our focus is on the initial phase
at the onset of inflation, it is possible
the effect of such damping terms also may
be applicable to the initial condition problem
in some supercooled inflation models.

\subsection{Radiation Component}
\label{radiation_component}
So far we only considered the inflaton field system, but
in addition the universe could possess some background component of
radiation energy density, $\rhorbg$.  To realize inflation, minimally
it requires the vacuum energy density $\rho_{\rm v}$
to dominate in the patch, $\rhorbg < \rho_{\rm v}$. 
If initially $\rhorbg$
is larger than $\rho_{\rm v}$, expansion of the universe will
red-shift the radiation. Thus provided the vacuum energy sustains itself
during this early period, inflation eventually will start.
This is the standard new and warm inflation pictures.  

One could imagine that in some small patch inside a larger 
causally created spatial region,
the inflaton field configuration is reasonably smooth and is sustaining a 
sizable vacuum energy of magnitude
\begin{equation} 
\rho_{\rm v} \approx \rhorbg.
\label{rvapprr} 
\end{equation} 
As the universe expands $\rhorbg$ decreases and provided
$\rho_{\rm v}$ sustains itself, inflation will begin in the patch.
However at this moment,  there could be 
large energy fluctuations in the radiation
bath or other fields that are just outside the putative inflationary
patch.  One would like to understand how probable it is for such a
situation to curtail the inflation that had started in the patch.

To answer this question based solely on causality considerations,
one should consider the case of a perturbation
starting on the patch boundary and moving across the patch at the speed
of light, and make the extreme assumption that as the perturbation
overruns 
regions of the inflation patch, those regions convert back
to being non-inflating.  The question is what minimal initial
inflationary patch size is needed so that its expansion
under inflation is faster than its contraction due to the
impinging perturbations.  {}For the case of flat geometry $\Omega=1$,
this question was addressed in \cite{Goldwirth:1992rj} 
and they concluded that
the patch size should be at least 3 times the Hubble radius in
order for inflation to succeed in enlarging the
patch.  
{}For a general $\Omega$,
we can address this  question
by evaluating 
the maximum distance a perturbation can travel 
in the patch between the initial
$t_i$ and final
$t_f$ time of inflation
\begin{equation}
  \label{eq:distance travelled}
  x_{\rm pert} = a(t) \int_{t_i}^{t_f} a^{-1} \, dt .
\label{xpert}
\end{equation}
Note that since $a$ is growing rapidly, there is negligible error
in taking $t_f \rightarrow \infty$ which implies
$x_{\rm pert}$ essentially is the event horizon.
We have examined Eq.\ (\ref{xpert}) for a variety of supercooled
and warm inflation models and generally find
$x_{\rm pert} \approx \xmas$.  

Thus in the most ideal case, an
inflationary patch smaller than the event horizon can not
be stable.  Here an important point of syntax should be noted,
that the generic size here is the event horizon and not the Hubble radius,
although for the flat case, $\Omega=1$, both are 
of the same order.
The above is the ideal bound based on causality.  However,
realistic dynamics also must be considered.

One case which 
corresponds to external perturbations entering
into an initially inflating patch is the case treated
in Subsect. IIIA of the mixing of high wavenumber modes.  
{}For the inflaton field interacting with itself or with other fields,
we argued above that if a dissipative term of the form $\Gamma {\dot \phi}$
is present in the inflaton effective evolution equation, then mode
mixing up to wave-numbers $k/a < \Gamma$ will not
occur within the time scale relevant to the initial condition
problem.  

In terms of the radiation bath, suppose a small vacuum 
dominated inflationary  regime
emerges, which is immersed inside a larger 
region containing radiation and gradient energy density.
Since the inflationary patch has negative pressure, the surrounding
radiation will flow into it.  The degree to which
the pressure differences are significant to this
process depends on the magnitudes
of the radiation, gradient and vacuum energies in both
regions as well as detailed dynamical 
considerations\cite{bkt,bubble}\footnote{The dynamics involves viscosity effects
at the interface between the inflation and background regions.
This viscosity is unrelated to the dissipative coefficient
$\Gamma$ in Sec. \ref{scalar_field}, which represents
the damping of the scalar field amplitude within the inflationary
patch.}.
Nevertheless, suppose the radiation tries to become uniform over
the entire region, including the patch.  
Since the patch started to inflate, it meant that initially
$\rho_{\rm v} \gtrsim \rhorbg$.  As time commences,
provided $\rho_{\rm v}$ remains constant, since expansion of the background
will decrease $\rhorbg$,  it means the amount of radiation
that flows into the inflationary patch will be less than $\rho_{\rm v}$.
Thus irrespective of whether the inflaton dynamics is supercooled
or warm, the patch should sustain a type of warm inflation due to
this influx of radiation from the regions that surround it. 

A thorough understanding of the initial
condition dynamics in presence of a background radiation component
is complicated.  To our knowledge, no quantitative analysis
has been done along these lines.  The review \cite{Goldwirth:1992rj}
only illustrated the problem with the example 
quoted above Eq.\ (\ref{xpert})
based on causality considerations but offered no dynamical 
examples with respect to a background radiation component.
Here we have offered one
scenario where the background radiation component  
should not prevent inflation from occurring.  In particular,
this example demonstrates that causality bounds 
on the rate at which radiation or other energy fluctuations
enter the patch are not the only aspect of this problem.
In addition, the magnitude of the entering
radiation must be adequately large to overwhelm the
vacuum energy.  We have offered arguments above that indicate this
latter requirement is not generic.  
 
\section{Dynamical Effects on Embedding}
\label{dynamical_effects_on_embedding}
As discussed in Sec. \ref{embedding_conditions}, a causally favourable
embedding requires the inflationary patch to start smaller than the
background MAS and then grow larger than it. A minimum condition for
this to happen is that the particle horizon of the
background eventually becomes greater than the background MAS.

For $p=\omega \rho$, the particle horizon is given by \cite{KT}
\begin{equation}
  \label{eq:particle_horizon}
  d(t) = \frac{1}{H} \int_0^1 \frac{dx}{\left[x^2(1-\Omega(t)) + \Omega(t) 
         x^{1-3w} \right]^{1/2}}.
\end{equation}
Comparing this equation with Eq.~(\ref{mascond}) it can be seen
that the particle horizon will only become greater than the MAS if
$\omega < 1/3$. This condition was also noted in \cite{DC} from a
slightly different perspective. There the patch was taken to start
larger than the background MAS.

It follows that our proposed embedding will not work for a pure
radiation background which has $\omega=1/3$. However, in general the
background will consist of radiation and an inhomogeneous scalar
field.  As discussed in Sec. \ref{dynamic_conditions}, the gradient
energy of the scalar field has $\omega = -1/3$ and the potential
energy has $\omega=-1$ while the kinetic energy has $\omega=1$. It
follows that if the background is gradient dominated our causal
embedding scheme will be viable, whereas our scheme fails if the
kinetic term dominates.  However, the kinetic energy of the scalar
field will be suppressed if the dissipative coefficient
$\Gamma$ of Sec. \ref{dynamic_conditions} is sufficiently large.
This is already required inside the putative inflation patch
in order to realize inflation.  Thus, it is not
unreasonable to expect $\Gamma$ in the background region
to be of the same magnitude. 

\section{Conclusion}
\label{conclusion}
This paper has investigated the initial condition
problem of inflation from the perspective both
of spacetime embedding and inflaton dynamics.  Our study
has highlighted two attributes of this problem which
have not been addressed in other works.  {}First, from
the perspective of spacetime embedding, we have observed that
the global geometry can play an important role in determining
the size of the initial inflationary patch that is consistent with
the weak energy condition.  Second, from the perspective of inflaton dynamics,
we have noted that a $\Gamma {\dot \phi}$
damping term could alleviate several
problems which traditionally have led to large scale homogeneity 
requirements before inflation.  

The purpose of this paper was to note for both these attributes, their
salient features with respect to the initial condition problem.
In the wake of this, several details emerge that must be understood.
Below, we will review the main result we found for both
attributes and then discuss the questions that must be addressed
in future work.  

{}For a causally generated patch a successful embedding can be achieved
if the patch does not contain an anti-trapped region.  We have shown
that the MAS size can be arbitrarily larger than the Hubble scale
provided $\Omega$ is made small enough.  So if the patch Hubble
horizon is taken as the minimum stable patch size, then the patch does
not have to contain an anti-trapped region if $\Omegainf < 1$.
This generalizes the analysis of \cite{Vachaspati:2000dy} which was
only for  $\Omega=1$. 
However, without the effects of damping, it
appears that the event horizon, not the Hubble horizon, 
is the minimum stable patch size. {}For
the de Sitter case one can see analytically that the event horizon is
equal to the MAS size regardless of $\Omega$ and numerical calculations
suggest the same is true for power law inflation. However, radiation
damping of perturbations could stabilize a patch smaller than the event
horizon. In this case an open geometry for the patch
would allow the patch not to contain an anti-trapped region and thus
allow a causal embedding in an expanding background which does not
violate the weak energy condition. 
Eventually the patch should develop a MAS within it, but by then it
could have expanded to be larger than the background MAS.

{}For the dynamics problem, with respect to the scalar field
the new consideration was the effect of a $\Gamma {\dot \phi}$
damping term.  We found that such a term could suppress
many of the effects from initial inhomogeneities of the inflaton,
which in studies traditionally done without this term
lead to important impediments to
entering the inflation regime.  It appears evident that
inclusion of such a damping term will
lead to qualitative differences in the
initial condition problem.  The most interesting
outcome is initial inflationary patches
smaller than the Hubble radius $1/{\rm H}$ may be able to inflate.

This paper examined the consequences of damping terms but did
not delve into their fundamental origin. 
{}For the cosmological setting,
such damping terms are typically associated with
systems involving a scalar field interacting
with fields of a radiation bath.  In this case, 
such damping terms have been
found in first principles calculations
for certain warm inflation models \cite{wifp},
although more work is needed along these lines. 
It is worth noting here that in the early stages of certain
supercooled inflaton scenarios where radiation is present, 
in particular new \cite{ni} and
thermal \cite{ti} inflation, a careful examination of the
dynamics may reveal damping terms similar to this.
Since the initial stages are the crucial period for the
initial condition problem, if further study supports the
importance of such damping terms, it may be useful
to better understand damping effects also in such
scenarios.

Since the most suggestive situation for the damping terms is
where in addition a radiation component is present in the universe,
in Subsect. IIIB we also studied the effects of this 
component on the initial condition problem.
Specifically we studied the case most suggestive
for the initial stages of new and warm inflation, 
where a small inflation patch is submerged inside a larger
radiation dominated spatial region.  Our main
observation has been that the minimal condition for
the patch to inflate is that its vacuum energy density must be larger than
the background radiation energy density.  Provided the vacuum
energy sustains itself, since expansion of the universe
will dilute the radiation energy density, it is not clear-cut
that the external radiation energy can act with sufficient magnitude
to impede inflation inside the patch.

We have shown that the gradient terms, due to the inhomogeneous scalar
field in the background space time, make it possible for the patch
boundary to overtake the background MAS. However the kinetic energy of
the scalar field must not be dominant for this to happen.  This can be
ensured by including a damping term in the scalar field equation.

\section{Acknowledgements}
We thank Robert Brandenberger, David Hochberg and Tanmay Vachaspati 
for helpful discussions. CG
also would like to thank his colleagues in the Relativity and Cosmology
Group for helpful discussions and in particular Roy Maartens for
detailed comments and he acknowledges the support of an ORS award. AB
acknowledges support of PPARC.




 






\end{document}